\relax
\documentclass[letterpaper]{article} 
\usepackage{aaai20}  
\usepackage{times}  
\usepackage{helvet} 
\usepackage{courier}  
\usepackage[hyphens]{url}  
\usepackage{graphicx} 
\usepackage{graphicx}  
\usepackage{amsmath}

\title{Sustained Online Amplification of COVID-19 Elites in the United States}

\author{
    Ryan J. Gallagher,\textsuperscript{\rm 1} 
    Larissa Doroshenko,\textsuperscript{\rm 2} 
    Sarah Shugars,\textsuperscript{\rm 1,3}\\
    \Large \textbf{David Lazer,\textsuperscript{\rm 1}
    and Brooke Foucault Welles\textsuperscript{1,2}}\\
    \textsuperscript{\rm 1}
        Network Science Institute, Northeastern University, Boston, MA 02115 \\
    \textsuperscript{\rm 2}
        Department of Communication Studies, Northeastern University, Boston, MA 02115 \\
    \textsuperscript{\rm 3}
        Center for Data Science, New York University, New York, NY 10011
}

\begin{document}



\maketitle

\begin{abstract}
    The ongoing, fluid nature of the COVID-19 pandemic requires individuals to regularly seek information about best health practices, local community spreading, and public health guidelines. In the absence of a unified response to the pandemic in the United States and clear, consistent directives from federal and local officials, people have used social media to collectively crowdsource COVID-19 elites, a small set of trusted COVID-19 information sources. We take a census of COVID-19 crowdsourced elites in the United States who have received sustained attention on Twitter during the pandemic. Using a mixed methods approach with a panel of Twitter users linked to public U.S. voter registration records, we find that journalists, media outlets, and political accounts have been consistently amplified around COVID-19, while epidemiologists, public health officials, and medical professionals make up only a small portion of all COVID-19 elites on Twitter. We show that COVID-19 elites vary considerably across demographic groups, and that there are notable racial, geographic, and political similarities and disparities between various groups and the demographics of their elites. With this variation in mind, we discuss the potential for using the disproportionate online voice of crowdsourced COVID-19 elites to equitably promote timely public health information and mitigate rampant misinformation.
\end{abstract}


\section{Introduction}

Local COVID-19 caseloads, community guidelines, national directives, international travel bans, testing locations, economic assistance, and scientific knowledge all change rapidly, sometimes on a daily or even hourly basis, as we learn more about COVID-19 as a disease and weather the pandemic. The fluidity of the ongoing public health crisis makes it difficult for individuals to find reliable information online, forcing them to sift through both high-quality expert information and life-threatening misinformation, particularly on social media platforms. As people individually engage with an imperfect assortment of local news media, epidemiological specialists, political pundits, public health officials, and financially-motivated hoaxes online, users collectively amplify a small set of disproportionately influential COVID-19 information sources. These accounts are key pillars in the online information ecosystem because they are the most capable of disseminating both critical health information and divisive anti-scientific opinions. Yet, despite their importance, we do not currently know who people have been consistently consulting on social media for information regarding the pandemic.

Here, we identify sustained sources of COVID-19 information in the United States on Twitter, a key vessel of information for citizens, journalists, and politicians alike. We draw on the concept of \emph{crowdsourced elites} \cite{papacharissi2012affective} to identify crowdsourced COVID-19 elites, Twitter accounts that are collectively deemed important sources of COVID-19 information. We introduce a measure that identifies sustained crowdsourced elites that are consistently amplified, and filters them from one-off viral accounts and those who do not consistently post about the pandemic. Using a panel of Twitter users matched to public U.S. voter registration files, we apply our measure of sustained amplification to enumerate crowdsourced COVID-19 elites, both nationally and across different demographic, geographic, and ideological subgroups of Twitter users in the United States. Nationally, we find that journalists, media outlets, and political accounts are most often crowdsourced as COVID-19 information sources, while epidemiologists, public health, and medical professionals make up only a small portion of all COVID-19 elites. As we show though, there is considerable variation in COVID-19 elites across demographics, which accentuates racial, political, and geographic homophily and disparities between COVID-19 elites and different subgroups. We discuss the implications of these patterns for leveraging crowdsourced COVID-19 elites to rapidly disseminate high-quality public health information, particularly among marginalized populations that are most devastated by COVID-19, so that we may counter hazardous misinformation online and mitigate the pandemic offline.


\section{Related Work}

\subsection{Networked Gatekeeping and Crowdsourced Elites}

The synergistic mix of journalists, media outlets, celebrities, and everyday people on Twitter gives the platform tremendous potential for disseminating information. In particular, Twitter has been critical during breaking news events \cite{hu2012breaking} and rapidly changing situations where traditional broadcast media outlets are not able to quickly adapt and provide fully-fledged reports \cite{grusin2010premediation}, such as social protests \cite{jackson2020hashtagactivism} and natural disasters \cite{pourebrahim2019understanding}. The fast pace of information on the platform makes it a timely, constantly updating alternative to mainstream media when they do not have on-the-ground or authoritative information \cite{grusin2010premediation}, or when they are blocked, restricted, or otherwise not trusted \cite{howard2010digital,papacharissi2009citizen}. The COVID-19 pandemic lies at the intersection of the reasons people seek information on Twitter. The uniqueness of the pandemic as a series of ongoing and constantly evolving public health crises, together with the absence of unanimous, authoritative guidance from federal and local health officials in the United States, sharp polarization in communication of elected officials \cite{green2020elusive}, and historically low trust in traditional media \cite{fink2019biggest}, has created confusion and distrust, prompting many to turn to social media for up-to-date, localized, and preferred COVID-19 information \cite{fisher2020with}. Although it is possible to find reliable health information, local directives, and other COVID-19 news on social media, the information is generally variable in quality, and can lead people to erroneous conclusions that are not grounded in any scientific facts about the virus \cite{fisher2020with}.

The specific COVID-19 information sources that people find online are curated through a collective, networked process. While in traditional news media journalists perform the function of gatekeeping, the process of selecting and verifying news information \cite{shoemaker2009gatekeeping}, social media shifts the balance of power in shaping news production \cite{chadwick2011political} and gives ordinary citizens the means to curate news \cite{schonfield2010twitter}. Individuals and audiences are able to amplify particular pieces of content and not others \cite{meraz2009many}, allowing them to act as gatekeepers on social networking sites. Through networked gatekeeping, social media users collectively \emph{crowdsource} information sources to prominence by horizontally filtering and promoting relevant content amongst each another \cite{papacharissi2012affective,meraz2013networked}. While many of these \emph{crowdsourced elites} can often match traditional elites \cite{hindman2008myth}, the particular affordances of Twitter allow for more permeability in who is ``elite,’’ compared with traditional media, and even older internet platforms. Specifically, during acute and rapidly evolving crises without authoritative information, local experts, activists, and otherwise ordinary individuals often emerge as crowdsourced elites \cite{jackson2016ferguson,pourebrahim2019understanding}. Their local, technical, and situational knowledge is amplified through the networked gatekeeping process, putting them alongside traditional elites, like broadcast media and elected officials.

\subsection{Sustained Crowdsourced Elites Across Communities}

Social media users invoke the networked gatekeeping process by sharing and amplifying accounts they see as relevant voices. In practice then, when the context is Twitter, crowdsourced elites are often identified by tallying the accounts with the most retweets \cite{meraz2013networked,jackson2016ferguson,stewart2017drawing}. This is an effective proxy for measuring which accounts were most visible, particularly during crises that emerge and dissipate relatively quickly. However, focusing only on retweet counts as a measure of the crowdsourcing mechanisms can be misleading in the context of a sustained crisis like the COVID-19 pandemic, a situation where constant engagement with the issue over several months or more is important. As a metric, the total number of reshares across all of an account’s posts conflates three types of accounts: those that had one viral post, those that had many low engagement posts, and those that consistently had a variety of their posts shared. Given that the pandemic is an ongoing crisis, it is critical to be able to distinguish crowdsourced COVID-19 elites that are consistently amplified from those that intermittently have their voice raised through viral posts but then just as quickly recede.

More generally though, constructing a single ranked list of crowdsourced elites according to any metric does not account for the ways in which different communities crowdsource different elites. This is particularly important because networked gatekeeping is susceptible to homophily effects, the tendency for individuals to interact with those similar to themselves \cite{mcpherson2001birds,wu2011says}. This can lead to a balkanization of information, where different subgroups collectively amplify very different information sources and crowdsource very different elites \cite{conover2011political,foucault2019battle}. On its own, this is not necessarily a bad thing, but, methodologically, if we only produce a single, overall ranking of crowdsourced elites and do not account for homophilous groups, then we may overestimate the importance of general elites for key demographic, geographic, and ideological communities; or, alternatively, we may underestimate the importance of community-specific elites for their particular constituencies. Research has consistently shown that behavioral change campaigns---including those designed to spread accurate information about things like health risks, treatment, and vaccination---are considerably more effective when trusted opinion leaders are enlisted to encourage their adoption and compliance \cite{valente1999accelerating,xu2014predicting}. Given that networked gatekeeping inherently results in the accounts that are collectively trusted to provide information, these community-specific crowdsourced elites may be able to trigger larger informational cascades \cite{cha2010measuring} and reach more people with critical public health information.


\section{Data and Methods}

We use a mixed methods approach for identifying the sustained amplification of COVID-19 elites during the first three months of the pandemic in the United States. We draw on a panel of Twitter users matched to public U.S. voting registration records, and identify COVID-19 content through a keywords-based approach. We then introduce a measure of sustained amplification, and rank accounts retweeted by our panel according to their consistent presence in COVID-19 discussions. Having identified COVID-19 crowdsourced elites across a number of demographics, we undertake a qualitative hand coding process and annotate them according to their type of account, race, gender, political affiliation, and geographic locality. We use those labeled accounts to map who people have been amplifying during the pandemic, and compare differences in crowdsourced elites across demographics.

\subsection{Twitter Panel of Registered Voters}

To focus our analysis on American Twitter users and how they amplify COVID-19 content, we use a panel of Twitter users linked with public U.S. voter registration records. The panel, developed in prior work
\cite{grinberg2019fake}, 
was constructed by first starting with a 10\% sample of tweets collected from the Twitter Decahose. From this sample, accounts were extracted if they reported a U.S. location in their profile and included a full name in their account or handle. These accounts were then joined with public voter registration files---assembled by the data vendor TargetSmart---by matching unique entries according to both name and location. For example, if Jane Doe was the only one with her name in Boston, Massachusetts, then her account was linked to the corresponding entry in the voter file. On the other hand, if there were multiple individuals with the name John Doe in Boston, none of them were included. In total, the matching process resulted in a Twitter panel of about 1.5 million registered U.S. voters, estimated to be about 3\% of all adult Twitter users in the U.S \cite{perrin2019share}.

Using the panel of Twitter users matched to U.S. voter registration records has several benefits over other methodological approaches. First, we can be quite confident that we are studying the online behavior of real people, and not that of media outlets, organizations, social bots, and other non-human entities \cite{ferrara2016rise,gorwa2020unpacking}. Second, as we describe momentarily, the panel makes it possible to study demographic variation in Twitter behavior, which is not otherwise possible using the Twitter API because Twitter does not solicit that information from its users. Finally, the size of the panel allows us to focus our analyses on demographic minorities and other marginalized populations while retaining a statistically reliable sample size \cite{foucault2014minorities}.

We make use of several demographic attributes from the panel data: state, race, gender, age, and political affiliation. The state of residence, age, and gender of our panel members come directly from the voter registration records. For race, only the nine states required by the Voting Rights Act consistently report race through voter registration records. For the remaining states, we use race as inferred by TargetSmart, who use a model based on the statistical distinctiveness of voters’ names with respect to their domicile in the United States. We recognize this is an imperfect measure, both statistically and ethically, though in aggregate the panel racial demographics tend to be consistent with survey estimates of Twitter demographics in the United States \cite{grinberg2019fake}. For party affiliation, we use TargetSmart’s estimate of party score, which is a statistical estimate based on a number of variables, including, for example, registered political party, number of votes in Republican and Democrat primaries, and FEC contributions. The score, which is more consistent across states then mere party registration, falls within an interval from 0 to 100, indicating the probability that a registered voter supports the Democratic party. Following TargetSmart’s recommended guidance, we classify any individual with a score less than 35 as a Republican and more than 65 as a Democrat.

Ethically, the linking of Twitter data to voter registration records warrants some concern. While we use public U.S. voter registration records and names and locations that were clearly and publicly disclosed on Twitter, many Twitter users likely do not anticipate having their information linked across these datasets \cite{fiesler2018participant,sloan2020linking} and their explicit consent was not obtained to do so. To defend against misuse of the data and any harm that may come to panel members from linking their Twitter profiles to their voter registration records, we store and analyze all data on secure servers with multiple checkpoints of restricted access. In addition, we do not include any private accounts in our analyses, and only report aggregate trends across the panel. Finally, the construction and use of the Twitter panel was approved by our Institutional Review Board. We believe these safeguards significantly mitigate the potential risks of linking the data. Given our goal of identifying critical information sources among Americans during the unprecedented public health crisis of the COVID-19 pandemic, particularly so that we may more equitably ensure minorities and marginalized populations have access to high-quality information, and that we implemented strict protections so that the linked data will not be abused, we believe that using the Twitter panel as we do here is warranted and ethically justified.

\subsection{COVID-19 Tweet Identification}

Using the Twitter panel as our source of tweets, we identify COVID-19-related content through a keyword-based approach. We compiled a multilingual keyword list starting from three sources: the keyword list for the COVID-19 Twitter Dataset assembled by Chen, Lerman, and Ferrara \shortcite{chen2020tracking}, the keyword list from Green et. al’s study on elite polarization around COVID-19 \shortcite{green2020elusive}, and the keyword filters for Twitter’s official COVID-19 streaming endpoint as of May 13th, 2020.\footnote{\url{https://developer.twitter.com/en/docs/labs/covid19-stream/filtering-rules}} We then augmented those lists with additional terms to expand their coverage across COVID-19 related topics and misinformation, and removed select words from the lists that were likely to produce false positives (e.g. ``china’’). In total, our list consists of 909 keywords, including, for example, terms explicitly mentioning the pandemic (``covid-19’’, ``coronavirus”), phrases referencing societal responses (``social distancing’’, ``flatten the curve’’, ``stayathomechallenge’’), names for different types of masks (``n95’’, ``surgical mask’’, ``face covering’’), and hashtags associated with misinformation (``plandemic’’, ``faucifraud’’, ``arrestbillgates2020’’). The full keyword list will be available as an online supplement.

We searched for these keywords in every tweet posted by our panel members from January 1st, 2020 to June 1st, 2020. We labeled a tweet as COVID-related if at least one keyword matched the tweet text, including any quoted text, any hashtags, or any substring of any URL included in the tweet. Finally, we identified COVID-related URLs and expanded our final tweet dataset to include any tweet using at least one of those URLs. A URL was identified as being COVID-related, even if it did not contain any keywords itself, if it was used with our keywords at least 100 times and at least 20\% of its use was with COVID keywords. Upon manual inspection, we found that these two filters identified a number of COVID-related URLs that were not otherwise labeled as such, while returning a minimal number of false positives. Overall, our tweet collection strategy resulted in 19,964,955 tweets posted by 483,683 panel members, where 14,233,300 of those tweets are retweets. Over 96\% of the tweets were posted after March 1st, 2020, when COVID-19 first became a focal point of conversation in the United States. The small proportion of tweets identified prior to the pandemic suggests that our keyword list has a generally low false positive rate.

\subsection{Sustained Amplification}

Shares, and retweets on Twitter specifically, are a canonical measure of amplification on social media platforms. During emergent discussions that rapidly coalesce around particular offline events, hashtags, and news stories \cite{meraz2013networked,jackson2016ferguson,stewart2017drawing}, counting the total number of retweets that different accounts received is often an effective way of identifying crowdsourced elites, those who commanded the most attention in the online conversation. However, the count of retweets is not necessarily a measure of \emph{consistent} amplification, particularly during an extended and ongoing crisis like the COVID-19 pandemic. If we are interested in who is receiving sustained attention online during the pandemic, only counting retweets risks over-representing users who had a small number of very popular tweets about COVID-19, but who were otherwise not at the backbone of COVID-19 conversations, and those who authored a very large number of tweets, but who received very few interactions on any one of them. 

To balance between these extremes and identify \emph{sustained} amplification around COVID-19, we define a COVID-19 retweet $h$-index. The $h$-index, originally proposed for measuring academic citations \cite{hirsch2005index}, assigns an index of $h$ to a user if that user has at least $h$ COVID-related tweets which have $h$ retweets each. For example, a user with 20 COVID-related tweets that each have at least 20 retweets will have an $h$-index of 20, while a user with either a single viral tweet or many tweets each with only 1 retweet will only have an $h$-index of 1. This means that a user can only have a high COVID-19 $h$-index if they consistently contribute to COVID-19 conversations \emph{and} and receive attention for those contributions. The COVID-19 retweet $h$-index refines our typical measure of amplification so that it accounts for sustained voice over the course of a prolonged crisis.

\subsection{COVID-19 Crowdsourced Elites}

For every account retweeted by one of our Twitter panel members, we calculate both the total number of COVID-19 related retweets received by the panel and their COVID-19 $h$-index. We then rank all the accounts by each of these measures, yielding two ranked lists from which we extract the top 0.1\% of accounts. We do this both across the entire national panel and for each demographic of age, race, gender, state of residence, and political affiliation. For example, we identify the COVID-19 elites crowdsourced by residents of Massachusetts by calculating the $h$-index according to retweets from just panel members in Massachusetts. For demographics where the top 0.1\% resulted in less than 50 accounts, we include all 50 of the top accounts, beyond just the top 0.1\%. Across both rankings of retweets and $h$-index and all demographics, this ranking and aggregation process yielded 1,478 unique accounts: 1,245 accounts are crowdsourced elites according to total retweets, and 1,098 are sustained crowdsourced elites according to the $h$-index.

To compare COVID-19 crowdsourced elites across different measures of amplification and demographics, we then conducted a qualitative analysis to characterize them. We first devised a classification scheme of potential COVID-19 information sources: public health officials, medical professionals, epidemiologists, other public intellectuals, elected officials, public services (police, city offices, etc.), other political accounts, media outlets, journalists, religious leaders, entertainers, athletes, other organizations, other accounts, and accounts that do not exist anymore. We formalized this typology in a codebook that also specified codes and instructions for attributing demographic and professional information to the crowdsourced elites, including their gender, race, political affiliation, and geographic locality. Three authors and five research assistants were trained on the codebook and instructed to use publicly-available information to research the COVID-19 crowdsourced elites, identify the relevant attributes, and classify them according to the information source typology. The codebook was updated iteratively based on feedback from the coding team, and all data were re-coded following codebook updates. Any ambiguous cases, as well as all cases marked ``other’’ were reviewed by at least two coders prior to finalizing a code. Average pairwise agreement across all coders calculated on a 5\% sample of the data at the conclusion of coding ranged from 85\% on codes with more categories (geographic location, account type) to over 90\% on codes with fewer categories (gender, race, political ideology). The lowest pairwise agreement (account type) was 85.714\% and the highest pairwise agreement (gender) was 97.143\%.


\section{Results}

\begin{figure*}[ht!]
    \centering
    \includegraphics[scale=.5]{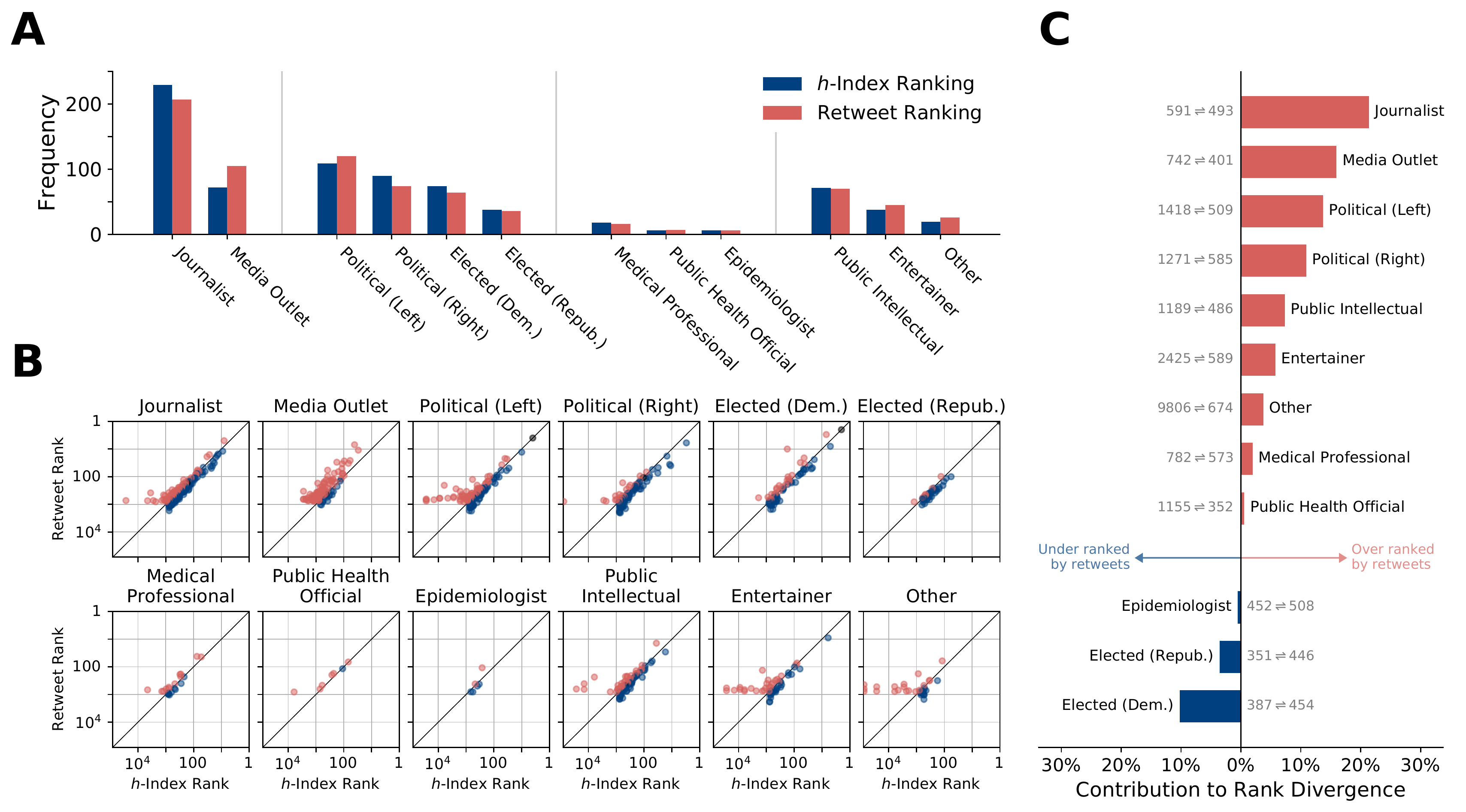}
    \caption{\textbf{A)} The distribution of COVID-19 crowdsourced elite account types, according to elites identified by counting total retweets (red) and calculating their retweet $h$-index (blue, see Data and Methods for details). \textbf{B)} Comparisons of accounts when ranking by their total number of retweets and $h$-index. Accounts above the line of equality in red are ranked higher by their total number of retweets, while accounts below the line of equality in blue are ranked higher by their $h$-index. \textbf{C)} Percent contribution of each type of elite account to the overall rank divergence \cite{dodds2020allotaxonometry}. All contributions are positive, but bars are directed according to whether the average rank of accounts of a particular type are higher according to total retweets (red, to the right) or $h$-index (blue, to the left). The annotations next to each bar show the average $h$-index rank (left) versus the average total retweets rank (right).}
    \label{fig:sustained-amplification}
\end{figure*}

\subsection{Sustained Amplification of COVID-19 Elites}

We first show that sustained amplification differs from raw amplification, as measured through total shares, and that raw amplification consistently overranks viral accounts relative to elites that consistently makeup the backbone of COVID-19 conversations. Using retweets from our national Twitter panel, we rank accounts according to both the total number of retweets they received on COVID-19 content and their COVID-19 retweet $h$-index (see Data and Methods for details). We take the top 0.1\% of both rankings as the crowdsourced elites, resulting in 813 elites for each ranking. All together, this yields 949 unique national elites, where 677 of them (83.3\%) are shared between the two rankings. President Donald J. Trump is ranked first according to both his total retweets and $h$-index ($n_{\text{RT}} = 92,536$, $h = 204$), followed by former vice president and current presidential candidate Joe Biden ($n_{\text{RT}} = 84,485$, $h=171$). The average elite ranked by their $h$-index received 6,869 retweets and had an $h$-index of 32, while the average elite ranked by their total retweets received 7,087 retweets and had an $h$-index of 30.

In Figure~\ref{fig:sustained-amplification}A, we show the distribution of account types among both the retweet and $h$-index rankings, as identified by our hand labeling. Regardless of the ranking, we see clear trends in what types of accounts have been amplified around COVID-19. Journalists and media outlets command a significant portion of the COVID-19 discussion, followed by political accounts on both the left and right. Notably across both rankings, medical professionals, public health officials, and epidemiologists make up only a small portion of all crowdsourced elites (about 3\%). Overall, online discussions of COVID-19 in the United States have been led by journalistic media and assorted political accounts, with less centering of accounts that may have more technical and frontline knowledge.

\begin{figure*}[!ht]
    \centering
    \includegraphics[scale=.46]{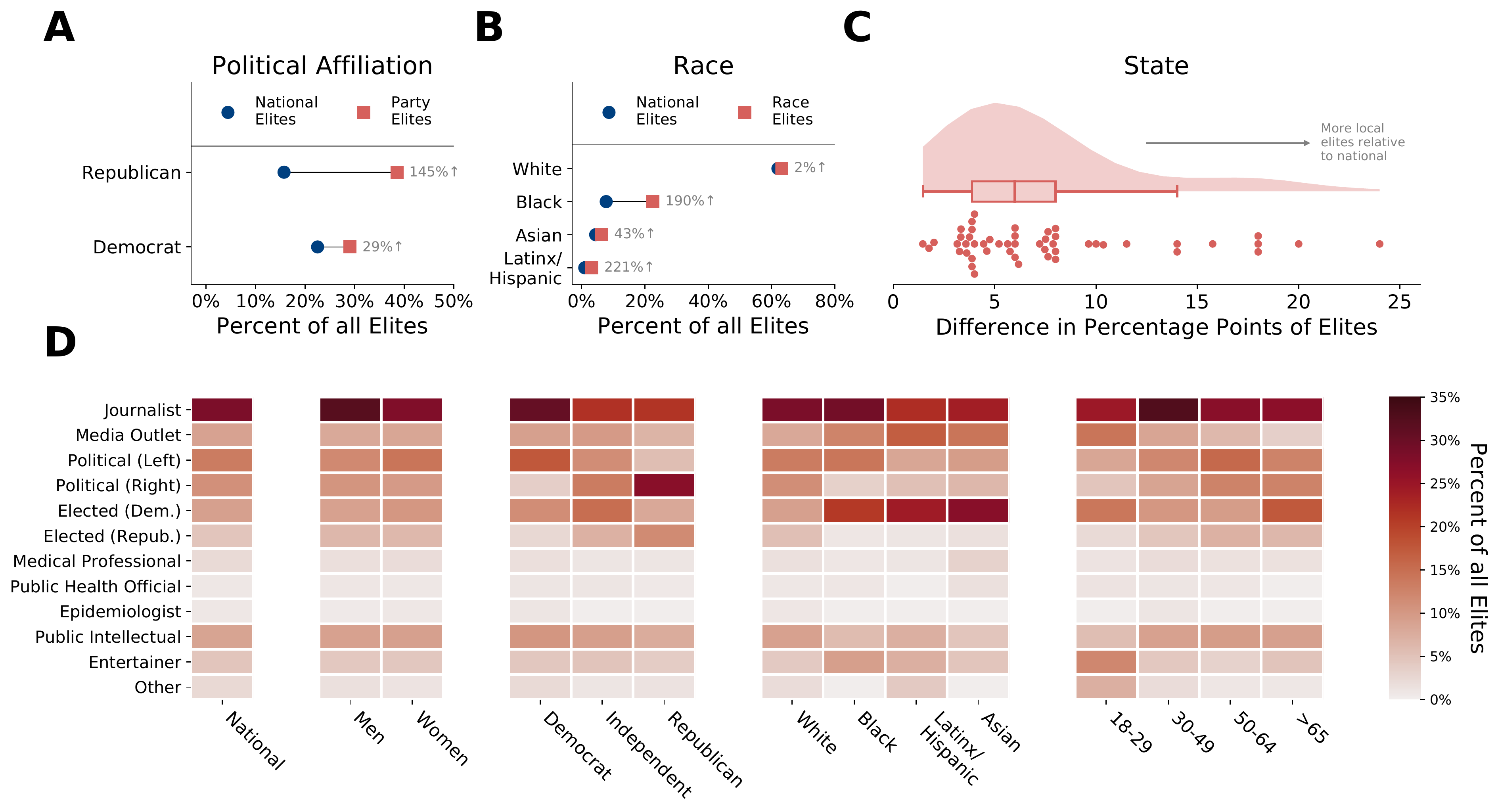}
    \caption{\textbf{A)} The political affiliation of elites both nationally and within-party (e.g., the percent of Democrats among nationally crowdsourced elites and among Democratic crowdsourced elites). \textbf{B)} The racial composition of elites both nationally and by particular racial categories (e.g., the percent of Black people among nationally crowdsourced elites and among Black crowdsourced elites). In \textbf{A)} and \textbf{B)}, annotations indicate the percent increase over the national percentage. \textbf{C)} Distribution of difference between within-state and national percentage of elites from a given state. Larger values indicate that more accounts from that state are crowdsourced locally than nationally. \textbf{D)} The distribution of COVID-19 elite account types nationally and across different demographics. Darker colors indicate relatively more accounts of that type crowdsourced as elites by that demographic.}
    \label{fig:variation}
\end{figure*}

Compared to the retweet $h$-index, which accounts for the sustained nature of the pandemic, raw totals of retweets overestimate the prevalence of media outlets among COVID-19 elites, and underestimate the prevalence of journalists. Total retweets also overestimate the number of elites that are political accounts on the left, underestimate the consistency of attention given to right-wing political accounts and, to a lesser extent, underestimate the sustained amplification of elected Democratic officials. Further though, the total retweets and $h$-index not only estimate different prevalences for these accounts, but rank them differently as well. In Figure~\ref{fig:sustained-amplification}B, we compare the relative rank of each crowdsourced elite according to their total retweets and $h$-index. A number of the plots, including those for journalists, media outlets, left-wing and right-wing political accounts, public intellectuals, and entertainers exhibit a red banding effect to the left, which indicates accounts that have high rank according to total retweets, but low rank according to $h$-index. These are the accounts that are particularly over ranked by only counting retweets: they may have a few tweets that garnered many retweets, but they do not consistently contribute to the COVID-19 discourse.

We quantify these overrankings more precisely using the rank divergence \cite{dodds2020allotaxonometry}. The rank divergence measures how differently the total retweets and $h$-index rank each account while, unlike other rank correlation measures, accounting for whether those shifts are across high, important ranks or lower, more peripheral ranks. The individual rank divergences are then aggregated together into the total rank divergence of each account type. We display these aggregate divergences in Figure~\ref{fig:sustained-amplification}, and annotate them with the average $h$-index and total retweet ranks of all accounts in each class. Both the rank divergences and differences in average ranks show how momentarily popular accounts can skew our understanding of who has been consistently crowdsourced as an elite during the pandemic. For example, the three panels of Figure~\ref{fig:sustained-amplification} together indicate that the prevalence of politically left accounts among COVID-19 elites is overestimated exactly because total retweets overweight the importance of one-off viral accounts that do not receive sustained amplification. Overall, journalists and media outlets contribute the most to the divergence between the retweet and $h$-index rankings, partly because of how they rank the accounts differently, but also because of the abundance of those types of accounts. On the other hand, while medical professionals and public health officials have significant differences in their average ranks between the two rankings (see annotations of Figure~\ref{fig:sustained-amplification}C), there are significantly fewer of these accounts, and so their overall contribution to the rank divergence is much lower.  In addition to overranking most account types because of viral outliers, total retweets also underrank the extent to which elected officials, both Republican and Democrat, have been consistently amplified around COVID-19.

\subsection{Demographic Crowdsourcing of COVID-19 Elites}

Having demonstrated our measure of sustained amplification, we now map the variation of COVID-19 elites among different demographic audiences. For each demographic of race, gender, age, state, and political affiliation in our Twitter panel, we calculate the COVID-19 retweet $h$-index for each account retweeted only by that demographic, and rank them accordingly. From those rankings we take the top 0.1\% of accounts to be the demographic’s COVID-19 crowdsourced elites. For demographics where the top 0.1\% of accounts yielded less than 50 total accounts, we took the top 50 rather than the top 0.1\%. We then compare the COVID-19 elites of each demographic to the national elites, those with the highest $h$-indices according to all retweets from members of our Twitter panel. There are 1,040 unique elites across all demographics, where 756 of those elites (72\%) are shared with the 813 national elites.

We first examine to what extent different demographic communities crowdsource elites from that community. For example, we measure what proportion of COVID-19 elites amplified by Republicans are themselves either Republican elected officials or right-wing political accounts, or what proportion of COVID-19 elites amplified by Black citizens are Black. We compare these to the same values calculated for the national panel, i.e. we compare the proportion of Black elites among Black panel members to the proportion of Black elites among the panel as a whole. The results across lines of political affiliation are shown in Figure~\ref{fig:variation}A, and the results across lines of race are shown in Figure~\ref{fig:variation}B. We did not find any preferential amplification of elites with respect to gender.

With respect to party, we find that both Democrats and Republicans respectively amplify more left and right wing COVID-19 elites than done so nationally. When looking specifically at political COVID-19 elites (those labeled as elected officials or otherwise political), rather than all elite account types, 82\% of Democrat’s political elites are left wing, and 73.9\% of Republican’s political elites are right wing (compared to 58.8\% and 41.2\% respectively among national political elites). With respect to race, we find that people of color are more likely to have COVID-19 elites that align with their racial demographic, compared to the national elites. Black panel members in particular have significantly more Black COVID-19 elites than the panel as a whole: while only about 8\% of COVID-19 elites nationally are Black, 23\% of COVID-19 accounts amplified by Black users are themselves Black. While the absolute number of Latinx/Hispanic and Asian COVID-19 elites is lower both nationally and for each demographic, Latinx/Hispanic panel members are about three times more likely to have a Latinx/Hispanic person among their elites, and Asian panel members are about 1.5 times more likely to crowdsource an Asian person as an elite. Generally, people of color are more likely to have COVID-19 elites that are of color: when looking at accounts that have a race (as opposed to organizations, for example), it is 71\% more likely for the elites of Black panel members to be non-white compared to those of white panel members, and 21\% and 12\% more likely for the elites of Latinx/Hispanic and Asian panel members to be non-white, respectively.

For each state, we find that there are relatively more local COVID-19 elites among the state’s elites than there are nationally. Figure~\ref{fig:variation}C shows the difference in percentage points between the percent of a state’s elites that are local journalists, media outlets, and elected officials and the percent of all national elites that are from that same state. The average state has 7.5 percentage points more local representation than is seen nationally, where 29 states have no local representation at all among national COVID-19 elites. Even states that are already highly represented among national elites, including California (2.5\% of all national elites) and New York (1.9\% of all national elites), crowdsource relatively more local elites: 4.5\% of all Californian elites and 3.3\% of all New York elites are affiliated with California and New York respectively.

Finally, for each demographic, we look at the distribution of account types that they have crowdsourced as COVID-19 elites, shown in Figure~\ref{fig:variation}D. Consistent with the aggregate national results of Figure~\ref{fig:sustained-amplification}A, we see that all demographics have consistently amplified journalists for COVID-19 information. Also consistent with the national trends, all demographics have crowdsourced very few medical professionals, public health officials, and epidemiologists as their COVID-19 experts. There are varying trends around political accounts for different demographics. As mentioned, Democrats and Republicans amplify more left- and right-wing accounts as their elites respectively. By the account type breakdown though, we see that both political demographics, and especially Republicans, crowdsource relatively more assorted political accounts than elected officials. Non-white panel members crowdsource very few right-wing elites, and place particular emphasis on the accounts of elected Democratic officials, rather than just general left-wing accounts. Political accounts of all types are crowdsourced more as elites among higher age bins: 29\% of elites for panel members ages 18-29 are political, 36\% are for those ages 30-49, 44\% are for those ages 50-64, and almost 50\% are for those ages 65 and older. In contrast to this trend, 14\% of elites for those ages 18-29 are media outlets, while that number falls for those ages 30-49 (8\%), 50-64 (6\%), and older (3\%).


\section{Discussion}

\subsection{Scientific Expertise and COVID-19 Elites}

In this paper, we have taken a census of who Americans consistently turn to for COVID-19 information on Twitter, and how those information sources vary across demographic groups. We introduced a measure of sustained amplification, which consistently identifies and ranks who lies at the backbone of COVID-19 discussions online. Applying this measure to a panel of American Twitter users, we documented that people have primarily amplified journalists and media outlets as COVID-19 elites, followed by assorted political accounts across the political spectrum. Further, we have also shown that epidemiologists, public health officials, and medical professionals only make up a small fraction of all COVID-19 elites crowdsourced nationally and across demographics. Consistent with other work on COVID-19 discussions online \cite{green2020elusive,jiang2020political}, these results indicate that Americans are receiving a sizeable portion of their COVID-19 information from political sources, including elected officials and political pundits, which likely results from and contributes to a polarized information ecosystem in the United States and politicizes the pandemic. Moreover, even sources that are not expressly political, such as mainstream journalists, may be perceived to be political by those with strong partisan affiliations, further exacerbating the politicization of COVID-19 information \cite{iyengar2009red}. Scientists and medical professionals, the most widely trusted sources across the political spectrum \cite{funk2020trust}, are scarcely present among the elites discovered in our analysis. While some prior work has estimated that scientific experts and authorities have made up a notable portion of COVID-19 discussions \cite{gligoric2020experts}, our results here suggest that those individuals are not consistently leaders in spreading COVID-19 information.

While our measure of sustained amplification can identify accounts that are persistently referred to for COVID-19 information, there are still a number of factors it cannot disentangle. Evidence suggests that the U.S. public, across demographic and ideological lines, trusts medical professionals and scientists more than elected officials and journalists, and this trust is on the rise since the start of the pandemic \cite{funk2020trust}. Yet, our results suggest scientists and medical professionals are not playing a central role in driving conversations about COVID-19 on Twitter, relative to other groups. This research cannot say exactly why this is the case; it may be that epidemiologists, public health officials, and medical professionals do not have established followings to amplify their messages, their messaging itself is ineffective, or they do not post about the pandemic as much as other types of accounts. All of these explanations suggest different interventions: the information of scientists and public authorities may need to be amplified further, their public communication needs to be improved, or they need to engage with the public more on Twitter generally. Regardless, given that all of the major social media platforms made a statement in March 2020 that they were “elevating authoritative content’’ on their platforms in response to the pandemic \cite{shu2020facebook}, and that a number of political, entertainment, and other public intellectuals have had significant voice in COVID-19 discussions, it is concerning that epidemiological experts and public health authorities are not more prominent among crowdsourced COVID-19 elites. This suggests that social media platforms may need to go further to ensure that experts are regularly centered in COVID-19 discussions and their messages well received.

\subsection{Marginality and Misinformation}

If the relevant scientific and public authorities are not regularly consulted for COVID-19 information, then effective public health messaging interventions should focus on crowdsourced COVID-19 elites who already share a sustained portion of COVID-19 discussions online. Importantly, we show that these elites vary notably across demographic groups, and that across lines of race, geography, and political affiliation, people are more likely to amplify accounts that are relevant to their beliefs and background. In particular, we showed that people of color were more likely to share COVID-19 information from elites of color. People of color are disproportionately affected by the pandemic \cite{wrigley2020us,azar2020disparities}, and our results suggest that there is a need for more non-white experts and authorities to communicate with these marginalized populations. Scientific experts and public officials could, for example, engage minority and local elites to transmit reliable information about COVID-19 to their respective communities online. While these steps will not solve offline racial inequities regarding COVID-19, they may help mitigate them to some extent, especially if crowdsourced elites can be used to combat misinformation targeting these groups. Given that the United States’ atrocious scientific track record of unethically experimenting on people of color, and Black people specifically, has undermined their trust in American health officials and scientific experts, these populations are primed for particular types of COVID-19 misinformation that can have fatal outcomes \cite{collins2020canaries}. Leveraging COVID-19 crowdsourced elites to communicate public health messaging to these marginalized communities may be a critical component in combating these kinds of misinformation.

Furthermore, while our current analysis highlights the types of people and accounts who have been sustained crowdsourced elites during the COVID-19 pandemic, we have not yet looked at the quality of the information being shared by these accounts. In itself, it is not necessarily concerning that content from journalists and media outlets are elevated significantly more than content from epidemiologists, public health officials, and medical professionals. Indeed, if the latter do not already have an established social media following, they may need media platforms to make their expertise visible through more traditional means then retweets. However, media outlets vary widely in their commitment and ability to sharing high-quality, accurate information, and the World Health Organization and other experts have warned of an ``infodemic’’ which has plagued the COVID-19 pandemic \cite{world2020immunizing,donovan2020social,gallotti2020assessing}. Previous work has further found notable partisan and age differences in fake news sharing \cite{grinberg2019fake}, suggesting that we may see partisan differences in the quality of content shared by party-sourced elites here. In future work, we plan to evaluate the quality of information shared by elites, with a particular focus on the amplification of misinformation and information pollution. We will also look more closely at the content being shared by local health departments and public officials, assessing the degree to which this content reflects the most accurate information that was available at the time it was shared.

\subsection{Limitations}

We acknowledge that there are limitations to our study. With respect to our Twitter panel itself, we have used blackbox inferences of race and political affiliation provided by a data vendor. While addressing biases in these inferences is beyond the scope of the current work, it is important for future work to continue mapping COVID-19 elites among these demographics so that we may better understand where our estimates here are problematic. With respect to COVID-19 elites, we only label a small portion of all users retweeted by our panel members, the top 0.1\%. This may partly explain why epidemiologists, public health officials, and medical professionals do not appear highly among crowdsourced COVID-19 elites. With respect to information seeking behavior, we cannot say to what extent COVID-19 crowdsourced elites were amplified algorithmically. Further, our panel members certainly acquired information beyond just retweets, and sought information directly from external sources, some of which they may have linked to in their own tweets. More broadly, Twitter is just one platform in the larger information ecosystem, and it would be incredibly valuable to understand how information crowdsourcing on Twitter interacts with crowdsourcing on platforms like Facebook, Reddit, YouTube \cite{marchal2020covid19}, Google, Instagram, WhatsApp, and TikTok, and with traditional broadcast media.

\section{Conclusion}

People will continue to need relevant, timely, and trusted information about COVID-19 in order to safeguard against the disease and adapt to its social consequences. Our work begins to map the information ecosystem that has emerged in the wake of the pandemic, and identifies sustained COVID-19 crowdsourced elites, those who have been widely and consistently amplified around the pandemic. These elites are focal points in online communication networks because they have been given an authoritative platform with respect to COVID-19, and their voices are more likely to resonate with the audiences that crowdsourced them. By working with COVID-19 elites to promote scientifically informed best practices and guidelines, particularly among populations already devastated by the pandemic, and addressing those that pollute the information ecosystem with dangerous and divisive misinformation, we may be able to leverage the natural crowdsourcing potential of social media to more equitably promote the public’s health.


\section*{Acknowledgements}

We thank Andrea Barrios, Kristen Flaherty, Sagar Kumar, V Lange, and Adina Gitomer, who made this project possible with their labeling of COVID-19 Twitter elites. We also thank Alan Mislove, Christo Wilson, Kenneth Joseph, Nir Grinberg, and Stefan McCabe for their help maintaining the panel data. This work was partially funded by grants from Northeastern University, the US Army Research Office (award W911NF-18-1-0421), and the National Science Foundation (award 2026631). We are grateful to our sponsors for their support. Any opinions, findings, and conclusions or recommendations expressed in this material are those of the authors and do not necessarily reflect the views of our sponsors.

\bibliographystyle{aaai}
\bibliography{bibl}

\end{document}